\documentclass{PoS}
\usepackage{amsmath}
\usepackage{graphicx}%
\usepackage{amsfonts}%
\usepackage{amssymb}

\newcommand{\beq}{\begin{equation}}
\newcommand{\eeq}{\end{equation}}
\newcommand{\bea}{\begin{eqnarray}}
\newcommand{\eea}{\end{eqnarray}}

\def\taun{{\cal T}_N}

\def\tauncut{{\cal T}_N^{cut}}
\def\tauonecut{{\cal T}_1^{cut}}

\title{NNLO phenomenology using N-jettiness subtraction}

\ShortTitle{NNLO phenomenology using N-jettiness subtraction}

\author{\speaker{Radja Boughezal}
\thanks{We thank the organizers for the kind invitation and for the very stimulating conference.}\\
        Argonne National Laboratory\\
        E-mail: \email{rboughezal@anl.gov}}

\abstract{We discuss next-to-next-to-leading order QCD results for Higgs, W-boson and Z-boson production in association with a jet in hadronic collisions, obtained using the recently developed N-jettiness subtraction method.}

\FullConference{12th International Symposium on Radiative Corrections (Radcor 2015) and LoopFest XIV (Radiative Corrections for the LHC and Future Colliders)\\
		15-19 June, 2015\\
		UCLA Department of Physics \& Astronomy Los Angeles, USA}
		
\begin{document}

\section{Introduction}

Run I of the Large Hadron Collider (LHC)  was marked by the discovery and initial characterization of the Higgs boson. 
The comparison of Standard Model (SM) predictions with data from Run I of the LHC was limited by the statistical precision of the experimental data.  This will no longer be the case during Run II, and systematic errors will dominate.  The largest systematic error currently hindering our understanding of Higgs properties is the theoretical precision of the SM predictions.  
This is the case for the well-measured di-boson modes~\cite{Higgserrors} which dominate the overall signal-strength determination.  
The theoretical uncertainties must be reduced in order to sharpen our understanding of the mechanism of electroweak symmetry breaking in Nature. Calculations through next-to-next-to-leading order (NNLO) in perturbative QCD have become increasingly necessary to match the 
precision of LHC measurements. In particular improvements in both the overall production rate of the Higgs boson and in the modeling of its kinematic distributions are needed to match the expected experimental precision of Run II. 

NNLO calculations for scattering processes with final-state jets at hadron colliders possess a complex singularity structure. 
Partial results for $2 \to 2$  processes such as inclusive jet production~\cite{Ridder:2013mf}, Higgs+jet production~\cite{Boughezal:2013uia,Boughezal:2015dra} and Z+jet production~\cite{Ridder:2016rzm} are available (full results for Z+j using antenna subtraction will be available soon~\cite{Ridder:upcoming}). More recently complete results for Higgs+jet production~\cite{Boughezal:2015aha}, $W^{+}$+jet 
production~\cite{Boughezal:2015dva} and $Z$+jet production~\cite{Boughezal:2015ded} were achieved using N-jettiness subtraction method~\cite{Boughezal:2015dva,Boughezal:2015eha,Stewart:2010tn,Gaunt:2015pea}. The purpose of this contribution is to summarize the method and the results achieved for these processes.

\section{Description of N-jettiness subtraction}
\label{sec:jettinessDef}

We review here the salient features of the N-jettiness subtraction scheme for NNLO calculations, which was recently 
introduced in the context of the NNLO computation of $W^{+}$ boson and Higgs boson production in association with 
a jet~\cite{Boughezal:2015aha,Boughezal:2015dva,Boughezal:2015ded}.  We begin with the definition of $N$-jettiness$,\taun$, a global event shape variable designed to veto final-state jets~\cite{Stewart:2010tn}:
\begin{equation}
\label{eq:taudef}
{\cal T}_N = \sum_k \rm{min}_i \left\{ \frac{2 p_i \cdot q_k}{Q_i} \right \}.
\end{equation}
The subscript $N$ denotes the number of jets desired in the final state, and is an input to the measurement.  For the Higgs + jet,
$W^{+} $+ jet and $Z $+ jet processes considered here, we have $N=1$.  Values of ${\cal T}_1$ near zero indicate a final state containing a single narrow energy deposition, while larger values denote a final state containing two or more well-separated energy depositions.  
The $p_i$ are light-like vectors for each of the initial-state beams and final-state jets in the problem, while the $q_k$ denote the four-momenta of any final-state radiation.  The $Q_i$ are dimensionful variables that characterize the hardness of the beam-jets and final-state jets.  We set $Q_i = 2 E_i$, twice the energy of each jet.  
The cross section for $\taun$ less than some value $\tauncut$ can be expressed in the form~\cite{Stewart:2010pd,Stewart:2009yx}
\begin{equation} \label{eq:fact}
 \sigma(\taun < \tauncut)=  \int H \otimes B \otimes B \otimes S \otimes   \left[ \prod_n^{N} J_n \right] +\cdots .
\end{equation}
The function $H$ contains the virtual corrections to the process.  The beam function $B$ encodes the effect of radiation collinear to one of the two initial beams.  It can be written as a perturbative matching coefficient convoluted with a parton distribution function.  $S$ describes the soft radiation, while $J_n$ contains the radiation collinear to a final-state jet.  The ellipsis denotes power-suppressed terms which become negligible for $\taun \ll Q_i$. Each of these functions obeys a renormalization-group equation that allows logarithms of $\taun$ to be resummed.  If this expression is instead expanded to fixed-order in the strong coupling constant, it reproduces the cross section for low $\taun$. The derivation of this factorization theorem in the small-${\cal T}_N$ limit relies upon the machinery of Soft-Collinear Effective Theory~\cite{Bauer:2000ew}.

The basic idea behind N-jettiness subtraction is that $\taun$ fully captures the singularity structure of QCD amplitudes with final-state partons.  This allows us to calculate the NNLO corrections to processes such as Higgs + jet, $W^{+} $+ jet and Z + jet in the following way.  We divide the phase space according to whether $\taun$ is greater than or less than $\tauncut$.  For $\taun > \tauncut$ there are at least two hard partons in the final state, since all singularities are controlled by N-jettiness.  This region of phase space can therefore be obtained from, for example, a NLO calculation of Higgs production in association with two jets in the case where the born process is Higgs + jet.  Below $\tauncut$, the cross section is given by the factorization theorem of Eq.~(\ref{eq:fact}) expanded to second order in the strong coupling constant.  As long as $\tauncut$ is smaller than any other kinematic invariant in the problem, power corrections below the cutoff are unimportant.

All ingredients of Eq.~(\ref{eq:fact}) are known to the appropriate order to describe the low $\taun$ region through second order in the strong coupling constant.  The two-loop virtual corrections for the processes discussed here are known~\cite{Gehrmann:2011aa,Gehrmann:2011ab}.
The beam functions are known through NNLO~\cite{Gaunt:2014xga,Gaunt:2014cfa}, as are the jet functions~\cite{Becher:2006qw,Becher:2010pd} and soft function~\cite{Boughezal:2015eha}.  It is therefore possible to combine this information to provide the full NNLO calculation of Higgs + jet
and $W^{+}/Z $+ jet. 

A full NNLO calculation requires as well the high $\taun$ region above $\tauncut$.  However, a finite value of $\taun$ implies that there are actually 
$N+1$ resolved partons in the final state.  This is the crucial observation; $\taun$ completely describes the singularity structure of QCD amplitudes that contain $N$ final-state partons at leading order.  The high $\taun$ region of phase space is therefore described by a NLO calculation with $N+1$ jets.  We must choose $\tauncut$ much smaller than any other kinematical invariant in the problem in order to a
void power corrections to Eq.~(\ref{eq:fact}) below the cutoff.  

\section{Numerical Results} \label{sec:numerics}

We now present some numerical results for Higgs, $W^{+}$ and Z production in association with a jet. For validation checks of the results presented here we refer the reader to the detailed description in~\cite{Boughezal:2015aha,Boughezal:2015dva,Boughezal:2015ded}.  We focus on 8 TeV proton-proton collisions.  Jets are reconstructed using the anti-$k_T$ algorithm~\cite{Cacciari:2008gp} with $R=0.5$.  For the Higgs+jet process
we show results using the NNPDF~\cite{Ball:2012cx} parton distribution functions (PDFs), for the $W^{+}$+jet we use 
CT10 PDFs~\cite{Gao:2013xoa} while for Z + jet we use CT14 PDFs~\cite{Dulat:2015mca}.  We use the perturbative order of the PDFs that is consistent with the partonic cross section under consideration: LO PDFs with LO partonic cross sections, NLO PDFs with NLO partonic cross sections, and NNLO PDFs with NNLO partonic cross sections.  We set the renormalization and factorization scales equal to the mass of the Higgs boson, $\mu_R = \mu_F = m_H$ for Higgs+jet, $\mu=M_W$ for $W^{+}$+jet and $\mu=\mu_0=\sqrt{m_{ll}^2+\sum (p_{T}^{jet})^2}$  for Z + jet. For the latter scale choice the sum is over the transverse momenta of all final-state jets, and $m_{ll}$ is the invariant mass of the di-lepton pair arising from the $Z$-boson decay.  To estimate the residual theoretical error, we vary these scales simultaneously around the central value by a factor of two. We set the mass of the Higgs boson as $m_H = 125$ GeV. We impose the following cuts on the final-state jet: $p_T^{jet} > 30$ GeV, $|\eta_{jet}|<2.4$ for Higgs+jet and $|\eta_{jet}|<2.5$ for $W^{+}$+jet. For the Z + jet process we show a plot for the dependence of the ratio $\sigma_{NNLO}/\sigma_{NLO}$ 
on the power corrections as a function of $\tau_1^{cut}$ and refer the reader to the corresponding paper for more phenomenological studies~\cite{Boughezal:2015ded}.

\begin{figure}[!tbp]
  \centering
  \begin{minipage}[b]{0.47\textwidth}
    \includegraphics[width=\textwidth]{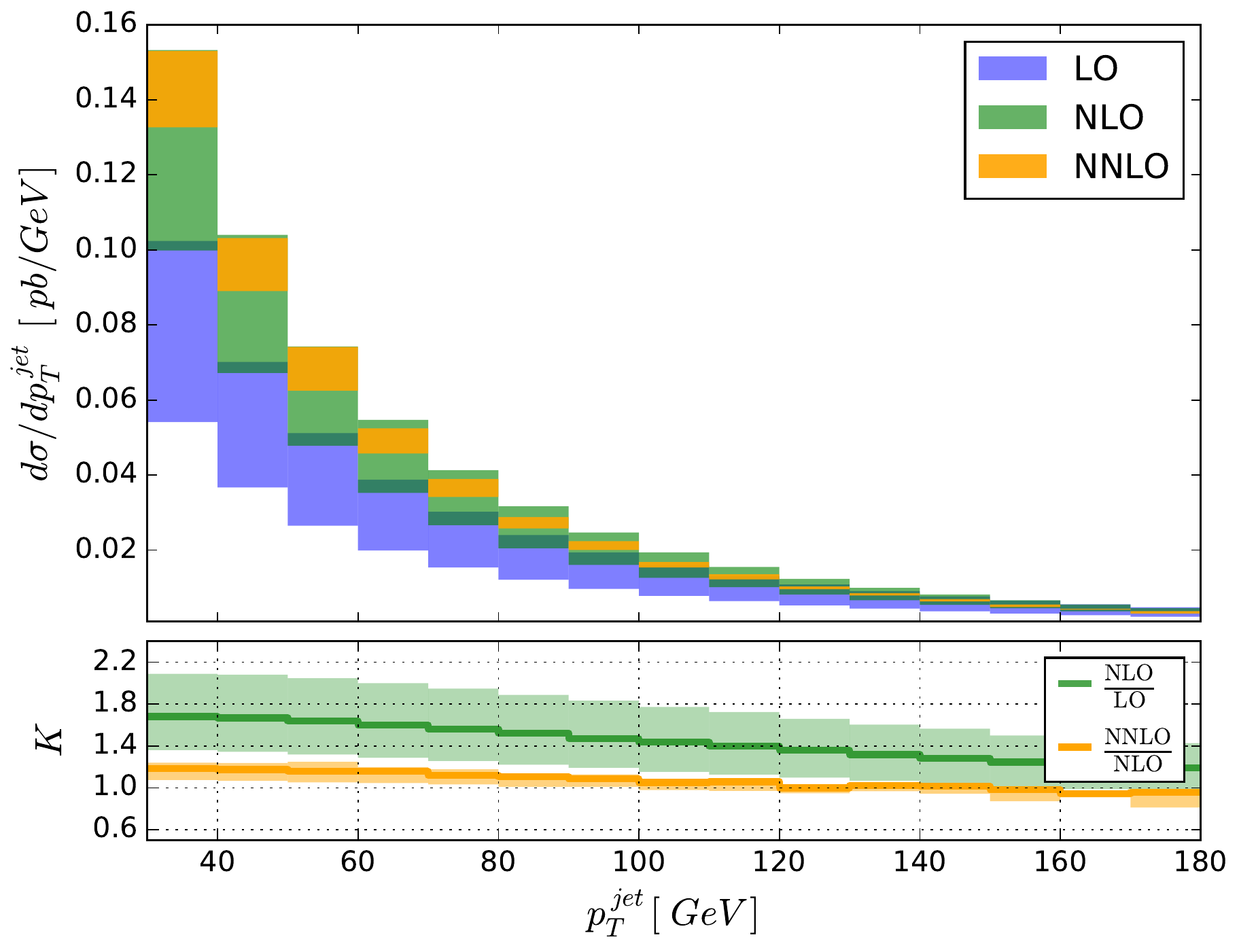}
    \caption{The transverse momentum of the leading jet for the Higgs+jet process at LO, NLO, and NNLO in the strong coupling constant. The lower inset shows the ratios of NLO over LO cross sections, and NNLO over NLO cross sections.  Both shaded regions in the upper panel and the lower inset indicate the scale-variation errors.} \label{fig:pTjet}
  \end{minipage}
  \hfill
  \begin{minipage}[b]{0.47\textwidth}
    \includegraphics[width=\textwidth]{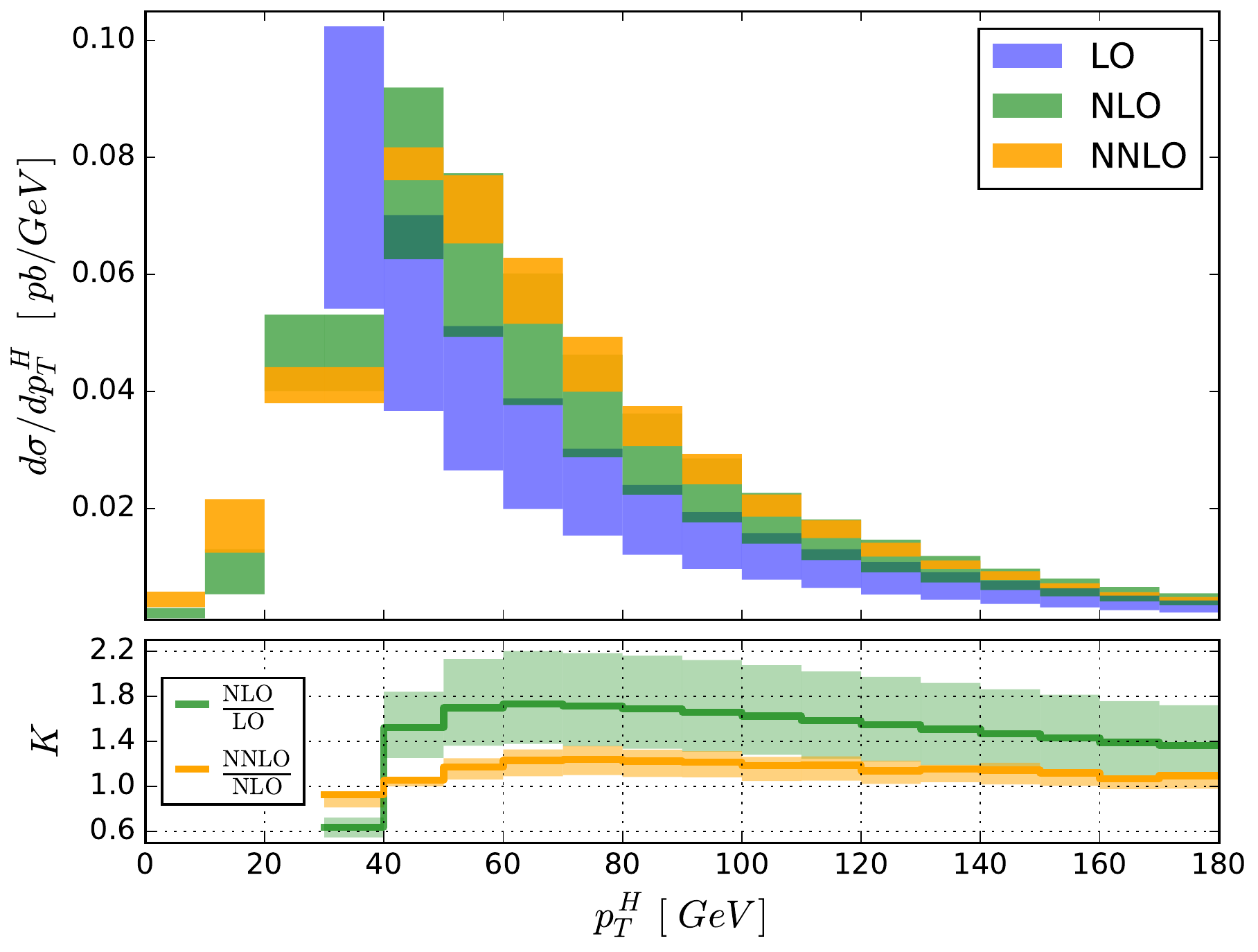}
    \caption{The transverse momentum of the Higgs boson at LO, NLO, and NNLO in the strong coupling constant. The lower inset shows the ratios of NLO over LO cross sections, and NNLO over NLO cross sections.  Both shaded regions in the upper panel and the lower inset indicate the scale-variation errors.} \label{fig:pThiggs}
  \end{minipage}
\end{figure}

We begin by showing few distributions in the Higgs plus jet production case.  In Fig.~\ref{fig:pTjet} we show the transverse momentum distribution of the leading jet. There is a shape dependence to the corrections, with the $K$-factor decreasing as $p_T^{jet}$ is increased. 
This trend is visible when going from LO to NLO in perturbation theory, and also when going from NLO to NNLO.  We note that the NNLO 
result is entirely contained within the NLO scale-variation band.  The shape dependence and magnitude of the NNLO corrections for the $p_T^{jet}$ distribution are in agreement with the results of Ref.~\cite{Boughezal:2015dra}, obtained using sector-improved residue subtraction 
scheme~\cite{Czakon:2010td,Boughezal:2011jf}. In Fig.~\ref{fig:pThiggs} we show the transverse momentum of the Higgs boson. The NLO corrections range from 40\% to 120\% near $p_{T}^H=60$ GeV, depending on the scale choice.  The magnitude of this correction decreases as the transverse momentum of the Higgs increases.  The NNLO corrections are more mild, reaching only 20\% at most for the central scale choice $\mu=m_H$.  They also decrease slightly as the transverse momentum of the Higgs increases.  The shape dependence and magnitude of the NNLO corrections for the $p_T^{H}$ distribution are in agreement with the results of Ref.~\cite{Boughezal:2015dra}.  We note that we have combined the two bins closest to the boundary $p_T^H=30$ GeV to avoid the well-known Sudakov shoulder effect~\cite{Catani:1997xc}.

\begin{figure}[!tbp]
  \centering
  \begin{minipage}[b]{0.47\textwidth}
    \includegraphics[width=\textwidth]{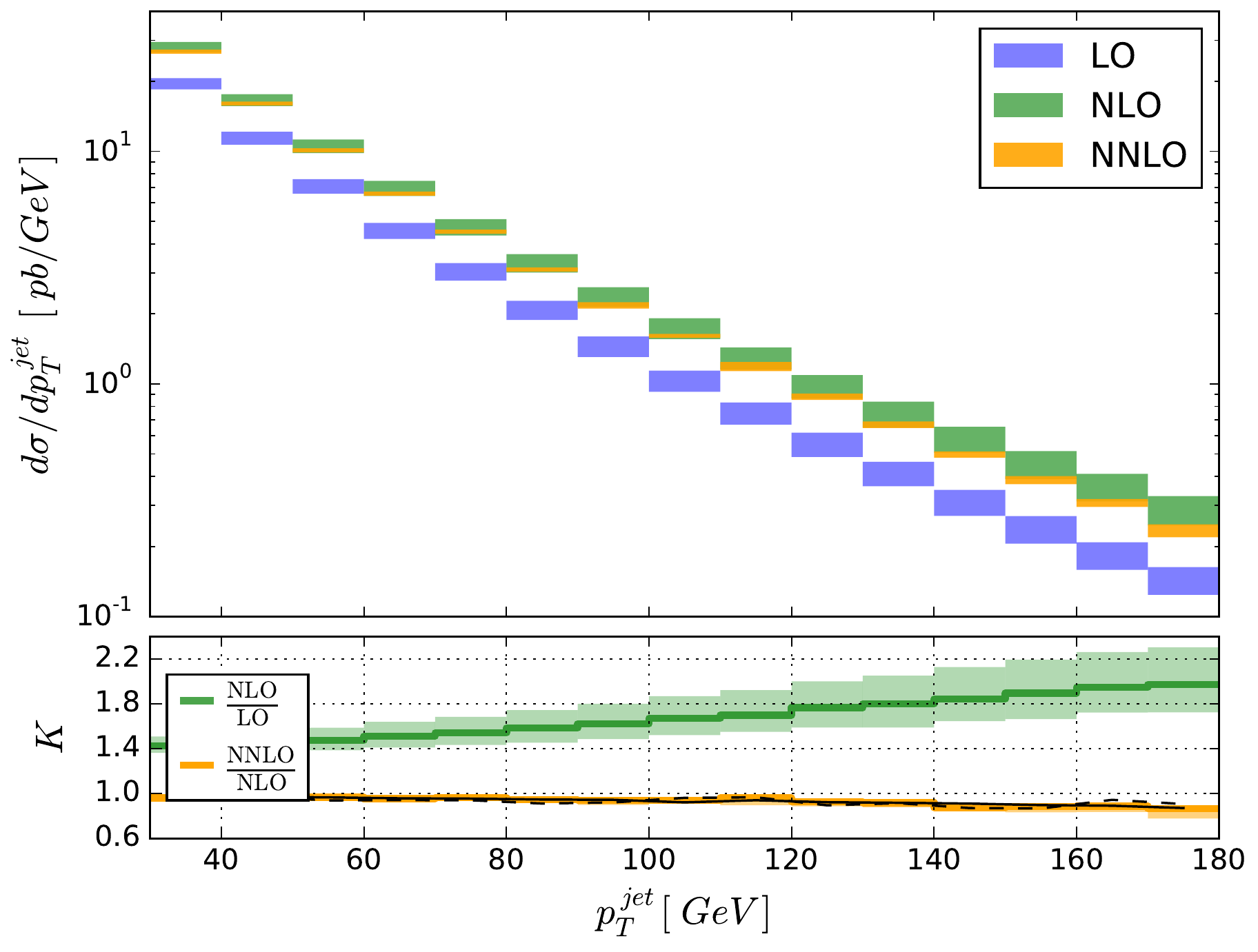}
    \caption{The transverse momentum spectrum of the leading jet for $W^{+}$ + jet at LO, NLO and NNLO in perturbation theory.  The bands indicate the estimated theoretical error. The lower inset shows the ratios of the NLO over the LO cross section, and the NNLO over the NLO cross section.  Both shaded regions in the upper panel and the lower inset indicate the scale-variation errors.  The dashed and solid black lines in the lower inset respectively show the distribution for $\tauonecut=0.05$ GeV and $\tauonecut=0.07$ GeV, for the scale choice $\mu=2 M_W$.} \label{fig:ptspectrum}
  \end{minipage}
  \hfill
  \begin{minipage}[b]{0.475\textwidth}
    \includegraphics[width=\textwidth]{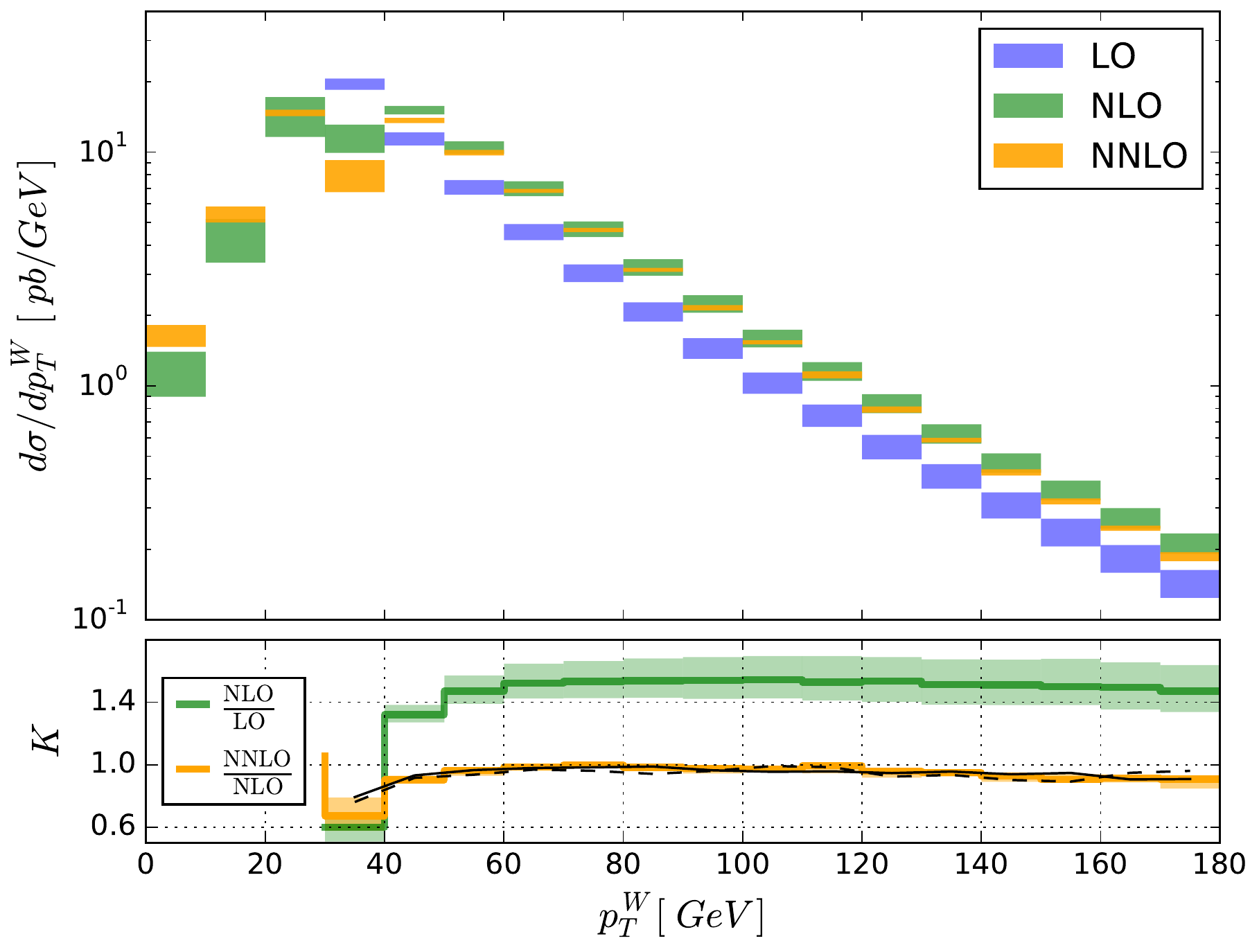}
    \caption{The transverse momentum spectrum of the $W^{+}$-boson at LO, NLO and NNLO in perturbation theory.  The bands indicate the estimated theoretical error. The lower inset shows the ratios of the NLO over the LO cross section, and the NNLO over the NLO cross section. Both shaded regions in the upper panel and the lower inset indicate the scale-variation errors.  The dashed and solid black lines in the lower inset respectively show the distribution for $\tauonecut=0.05$ GeV and $\tauonecut=0.07$ GeV, for the scale choice $\mu=2 M_W$.}  \label{fig:ptWspectrum}
  \end{minipage}
\end{figure}

In Fig.~\ref{fig:ptspectrum} we show the transverse momentum spectrum of the leading jet for $W^{+}$ + jet at LO, NLO and 
NNLO in perturbation theory.  The ratios of the NLO cross section over the LO result, as well as the NNLO cross section over the NLO one, 
are shown in the lower inset.  The shaded bands in the upper inset indicate the theoretical errors at each order estimated by varying 
the renormalization and factorization scales by a factor of two around their central value, as do the shaded regions in the lower inset.  In the
 lower inset we have shown the results for both $\tauncut=0.05$ GeV and $\tauncut=0.07$ GeV, for the scale choice 
 $\mu=2 M_W$, to demonstrate the $\tauncut$ independence in every bin studied.  The NLO corrections are large and positive for this scale choice, increasing the cross section by 40\% at $p_T^{jet}=40 $ GeV and by nearly a factor of two at $p_T^{jet}=180$ GeV.  The scale variation at NLO reaches approximately $\pm 20\%$ for $p_T^{jet}=180$ GeV.  The shift when going from NLO to NNLO is much more mild, giving only a percent-level decrease of the cross section that varies only slightly as $p_T^{jet}$ is increased.  The scale variation at NNLO is at the percent level and is nearly invisible on this plot.

The transverse momentum spectrum of the $W$-boson is shown in Fig.~\ref{fig:ptWspectrum}.  The NLO 
corrections are again 40\% for $p_T^W \geq 50$ GeV with a sizable scale dependence, while the NNLO corrections 
are flat in this region and decrease the cross section by a small amount.  The phase-space region 
$p_T^{W}<30$ GeV only opens up at NLO, leading to a different pattern of corrections for these transverse momentum values.  
The instability of the perturbative series in the bins closest to the boundary $p_T^W=30$ GeV is caused by the well-known Sudakov-shoulder 
effect~\cite{Catani:1997xc}.

\begin{figure}[h]
\centering
\includegraphics[width=3.3in]{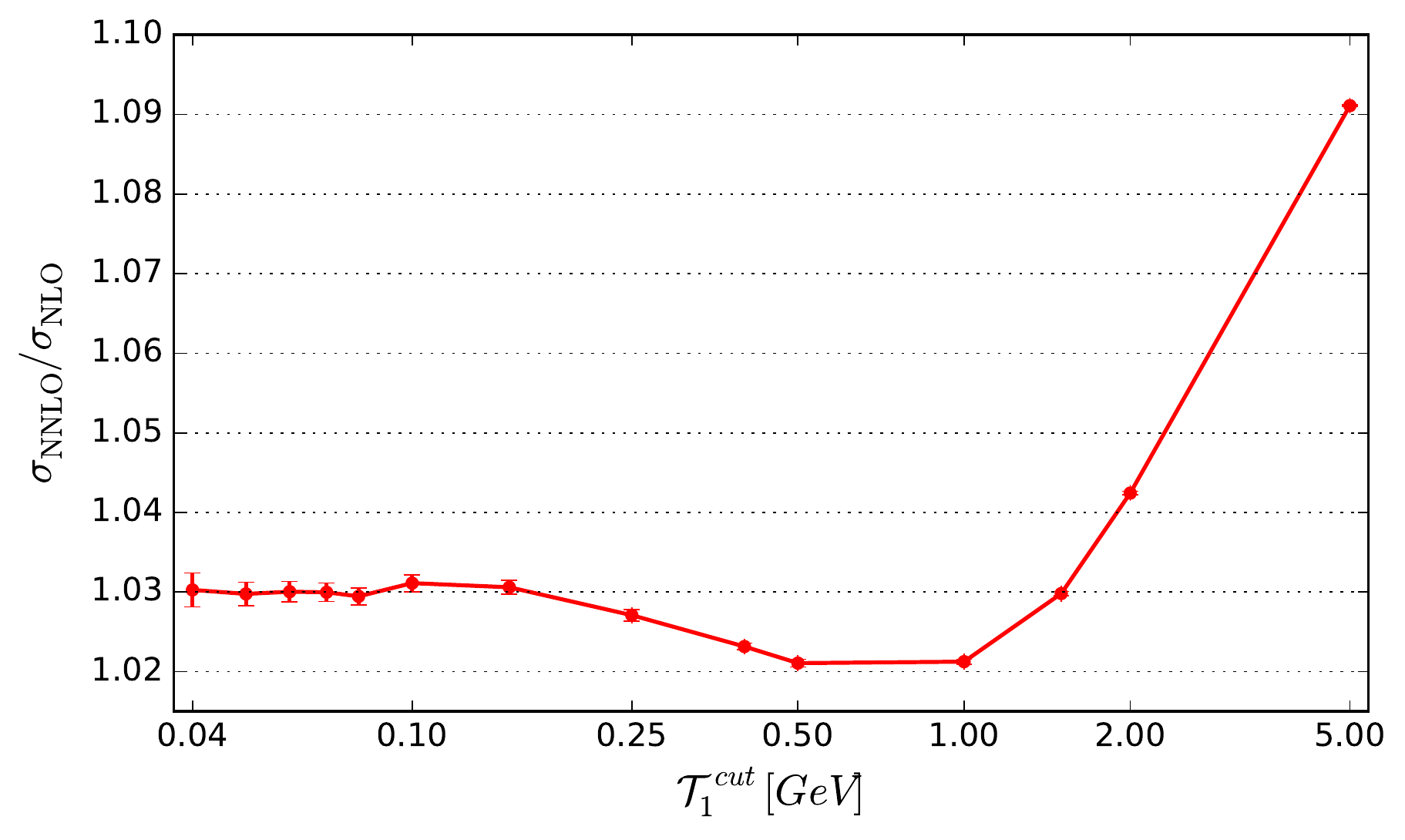}%
\caption{Plot of the NNLO cross section over the NLO result, $\sigma_{\text{NNLO}}/\sigma_{\text{NLO}}$, as a function of $\tauonecut$, for the scale choice $\mu= 2 \times \mu_0$.  The vertical bars accompanying each point indicate the integration errors.} \label{fig:taucheck}
\end{figure}

Finally, in Fig.~\ref{fig:taucheck} we show the dependence of the sum of the cross sections above and below the N-jettiness cutoff
$\tauncut$ and the effect of power corrections. The validation is done for the ratio $\sigma_{\text{NNLO}}/\sigma_{\text{NLO}}$ in
 13 TeV proton-proton collisions.  We have checked that the NLO cross section obtained with $N$-jettiness subtraction agrees with 
the result obtained with standard techniques.  These cross sections are obtained using CT14 PDFs at the same order in perturbation theory as the partonic cross section, and contain the following fiducial cuts on the leading final-state jet and the two leptons from CMS~\cite{Khachatryan:2014zya}: $p_T^{jet} > 30$ GeV, $|\eta_{jet}|<2.4$, $p_T^{l} > 20$ GeV, $|\eta_{l}|<2.4$ and $71\, \text{GeV} ~< m_{ll} ~< 111 \, \text{GeV}$. The ATLAS analysis is similar but with slightly different cuts.  We reconstruct jets using the anti-$k_T$ algorithm~\cite{Cacciari:2008gp} with $R=0.5$.  
A dynamical scale $\mu_0=\sqrt{m_{ll}^2+\sum p_{T}^{jet,2}}$ is chosen to describe this process, where the sum is over the transverse
 momenta of all final-state jets, and $m_{ll}$ the invariant mass of the di-lepton pair arising from the $Z$-boson decay. In this 
validation plot we have set the renormalization and factorization scales to $\mu_R = \mu_F = 2 \times \mu_0$; since the corrections are larger for this scale choice, it is easier to illustrate the important aspects of the $\tauonecut$ variation.

A few features can be seen in Fig.~\ref{fig:taucheck}.  First, in the region $\tauonecut < 0.2$ GeV the result becomes independent of the particular value of the cut chosen within the numerical errors.  The NNLO correction for $\mu = 2 \times \mu_0$ corresponds to a $+3\%$ shift in the cross section.  The plot makes clear that we have numerical control over the NNLO cross section to the per-mille level, completely sufficient for phenomenological predictions.  We observe an approximately linear dependence of $\sigma_{\text{NNLO}}$ on ${\rm ln}\left(\tauonecut\right)$ in the region $ 0.2 \, {\rm GeV} < \tauonecut < 0.5 \, {\rm GeV}$, indicating the onset of the neglected power corrections.  These power corrections have the form $(\taun / Q) \, {\rm ln}^n(\taun / Q)$, where $n \leq 3$ at NNLO~\cite{Gaunt:2015pea} and $Q$ is a hard scale such as $p_T^{jet}$.

\section{Conclusions}

We have presented in this proceedings the complete NNLO calculation of $W^{+}/Z$ + jet and Higgs boson production in association with a jet in hadronic collisions.  To perform this computation we have used a new subtraction scheme based on the $N$-jettiness event-shape variable  $\taun$.  
We will further study the phenomenological impact of our NNLO result in future work, including the prediction for the exclusive one-jet bin, where an intricate interplay between various sources of higher-order corrections was recently pointed out~\cite{Boughezal:2015oga}.
We will in addition use the full NNLO result to improve upon the resummation of jet-veto logarithms that occur when Higgs production is measured in exclusive jet bins.  Previous work has indicated that this resummation has an important effect in reducing the theoretical uncertainties that plague the predictions for exclusive jet multiplicities~\cite{Boughezal:2013oha}.

\end{document}